\documentclass[journal]{IEEEtran}

\usepackage{graphicx}
\usepackage{bm}
\usepackage{enumerate}
\usepackage{float}
\usepackage{multicol}
\usepackage{color}
\usepackage{url}
\usepackage{cite}
\usepackage{amsmath}
\usepackage{amsthm}
\usepackage{amssymb}
\usepackage{array}
\usepackage{multirow}
\usepackage[table,xcdraw]{xcolor}

\usepackage{balance}

\renewenvironment{IEEEbiography}[1]
  {\IEEEbiographynophoto{#1}}
  {\endIEEEbiographynophoto}

\ifCLASSOPTIONcompsoc
  \usepackage[caption=false,font=normalsize,labelfont=sf,textfont=sf]{subfig}
\else
  \usepackage[caption=false,font=footnotesize]{subfig}
\fi

\usepackage{acronym}

\PassOptionsToPackage{hyphens}{url}
\usepackage{hyperref}

\acrodef{AoA}{angle-of-arrival}
\acrodef{AoD}{angle-of-departure}
\acrodef{BS}{base station}
\acrodef{BF}{beamforming}
\acrodef{BPL}{building penetration loss}
\acrodef{bps}{bit-per-second}
\acrodef{RA}{receive antenna}
\acrodef{CP}{cooperative}
\acrodef{NCP}{non-cooperative}
\acrodef{PA}{predictor antenna}
\acrodef{CDF}{cumulative distribution function}
\acrodef{DFT}{discrete Fourier transform}
\acrodef{DU}{distributed unit}
\acrodef{DSRC}{dedicated short-range communication}
\acrodef{HARQ}{hybrid automatic repeat request}
\acrodef{PDF}{probability density function}
\acrodef{CU}{centralized unit}
\acrodef{c.u.}{channel use}
\acrodef{SNR}{signal-to-noise ratio}
\acrodef{SINR}{signal-to-interference-plus-noise ratio}
\acrodef{CSIT}{channel state information at the transmitter}
\acrodef{CSI}{channel state information}
\acrodef{bps}{bits per second}
\acrodef{npcu}{nats per channel use}
\acrodef{E2E}{end-to-end}
\acrodef{MRC}{maximum ratio combining}
\acrodef{MT}{mobile terminal}
\acrodef{MRT}{maximum ratio transmission}
\acrodef{OFDM}{orthogonal frequency division multiplexing}
\acrodef{LoS}{line-of-sight}
\acrodef{LiDAR}{light-detection-and-ranging}
\acrodef{LSCPA}{large-scale cooperative PA}
\acrodef{NLoS}{non-line-of-sight}
\acrodef{NLoSb}{non-line-of-sight, building}
\acrodef{NLoSbm}{non-line-of-sight, building, muted antenna}
\acrodef{NLoSv}{non-line-of-sight, vehicle}
\acrodef{MIMO}{multiple-input multiple-output}
\acrodef{MISO}{multiple-input single-output}
\acrodef{MRN}{moving relay node}
\acrodef{MR}{moving relay}
\acrodef{mmw}{millimeter wave}
\acrodef{UE}{user equipment}
\acrodef{4G}{fourth generation}
\acrodef{5G}{fifth generation}
\acrodef{LTE}{Long-Term Evolution}
\acrodef{LTE-A}{Long-Term Evolution-Advanced}
\acrodef{UL}{uplink}
\acrodef{DL}{downlink}
\acrodef{QoS}{quality-of-service}
 \acrodef{IoV}{Internet-of-Vehicle}
\acrodef{TDD}{time division duplex}
\acrodef{FDD}{frequency division duplex}
\acrodef{GPS}{Global Positioning System}
\acrodef{IAB}{integrated access and backhaul}
\acrodef{SIMO}{single-input-multiple-output}
\acrodef{SISO}{single-input-single-output}
\acrodef{ZF}{zero-forcing}
\acrodef{VPL}{vehicle penetration loss}
\acrodef{V2I}{vehicle-to-infrastructure}
\acrodef{V2X}{vehicle-to-anything}
\acrodef{3GPP}{the 3rd Generation Partnership Project}
\acrodef{URLLC}{ultra-reliable low-latency communication}
\acrodef{NR}{new radio}
\acrodef{DAPS}{dual active protocol stack}
\acrodef{OVT}{on-vehicle transceiver}
\acrodef{RIS}{reconfigurable intelligent surface}
\acrodef{EVM}{error vector magnitude}
\acrodef{LRBA}{low-complexity RIS-assisted blockage avoidance}
\acrodef{LSRPA}{large-scale based RIS pre-assignment}
\acrodef{gNB}{gNodeB}
\acrodef{DAPS}{dual active protocol stack}

\begin{document}
\captionsetup{belowskip=0pt,aboveskip=0pt}

\title{Dynamic Blockage Pre-Avoidance using  Reconfigurable Intelligent Surfaces}

\author{Hao~Guo,~\IEEEmembership{Graduate~Student~Member,~IEEE},
        Behrooz~Makki,~\IEEEmembership{Senior~Member,~IEEE},
        Magnus Åström,
        Mohamed-Slim Alouini,~\IEEEmembership{Fellow,~IEEE},
        and Tommy~Svensson,~\IEEEmembership{Senior~Member,~IEEE}    }

    
\maketitle

\begin{abstract}
Internet-of-vehicle (IoV) is a general concept referring to, e.g., autonomous drive based vehicle-to-everything (V2X) communications or moving relays. Here,  high rate and reliability demands call for advanced multi-antenna techniques and millimeter-wave (mmw) based communications. However, the  sensitivity of the mmw signals to blockage may limit the system performance, especially in highways/rural areas  with limited building reflectors/base station deployments and high-speed devices. To avoid the blockage, various techniques have been proposed among which reconfigurable intelligent surface (RIS) is a candidate. RIS, however, has been mainly of interest in stationary/low mobility scenarios, due to the associated  channel state information acquisition and beam management overhead as well as imperfect reflection. In this article, we study the potentials and challenges of RIS-assisted dynamic blockage avoidance in IoV networks. Particularly, by designing region-based RIS pre-selection as well as blockage prediction schemes, we show that RIS-assisted communication has the potential to boost the performance of IoV networks. However, there are still issues to be solved before RIS can be practically deployed in IoV networks.
\end{abstract}

\IEEEpeerreviewmaketitle

\section{Introduction}
With 4G and 5G, wireless networks have made great progress in serving stationary/low mobility devices. At high speeds, however, there are still various issues to be solved.   In order to connect everything at anytime and any place,  as one of the main objectives of 5G and beyond, \ac{V2X}, in general, \ac{IoV} \cite{guo2021iciot} communication plays an important role as the passengers on their vehicles expect the same \ac{QoS} as they experience at home.  Also, advanced vehicle technologies such as platooning, self driving and remote control require high-rate uninterrupted communications \cite{garcia2021CST}. For instance, according to 3GPP Use Case Groups \cite{3gppv2x},  remote driving may need reliability up to 99.999\% and latency down to 5 ms.

The intelligent vehicular communication systems were initially based on \ac{DSRC} standards  such as ITS-G5/\ac{DSRC}. Such technologies, which are based on  licensed bands and support speed range $\le 130$  km/h, reach less than 10 Mbps data rates with latency $\le 10$ ms.   \cite{guo2021iciot}. In \ac{LTE}, \ac{V2X} has been designed  for basic functionalities such as cooperative safety, road management, and telematics applications, requiring moderate data rates up to 100 Mbps. \ac{LTE}-based vehicular communication is based on spanning multiple bands in 450 MHz-4.99 GHz, with latency 100-200 ms and supported speeds up to 350 km/h. Also, it provides similar types of services as \ac{DSRC} to support transmission of basic road information, e.g., the location, velocity, and acceleration status of the vehicle \cite{garcia2021CST}.  

Compared to \ac{LTE}, 3GPP Release 16 proposes an evolutionary standard for \ac{V2X} using the 5G \ac{NR} air interface. Particularly, 5G \ac{NR} \ac{V2X} considers various levels of automation ranging from Level 0 with no automation to Level 5 with full automation, and the \ac{QoS} requirement increases with the automation level. Moreover, different use cases such as car platooning, advanced driving, extended sensors and remote driving use cases are defined each with its specific requirements. Specifically, in use cases such as extended sensor and remote driving, the vehicles can share their data obtained from own sensors with the surrounding infrastructure/vehicles. In this way, the perception of the environment is jointly improved for the network and it opens the opportunities for advanced cooperative schemes. Here, mobility management based on handover is the main focus where dual connectivity and UE-based handover, e.g., \ac{DAPS}-based handover, are defined. 5G \ac{NR} \ac{V2X} standardization  has continued in Release 17 for few enhancements (see \cite{garcia2021CST} for details).

With levels 3-5 of self-driving vehicles and the probable standardization of mobile \ac{IAB} in 3GPP Release  18, considerably higher rates may be required compared to those provided by \ac{LTE}. Here, following 5G \ac{NR}, there may be a need for using \ac{mmw} communications via small-cell deployments. Particularly, utilizing \ac{MIMO} techniques, beamforming and advanced resource allocation, as the inherent features of 5G \ac{NR}, \ac{mmw} communication has the potential to provide massive bandwidth and the data rates required for \ac{IoV} communications at typical vehicular speeds.

Although \ac{mmw} communication supports high data rates, it can be significantly affected by the penetration loss, along with severe path loss and beamforming mismatch. Utilizing small cells, deployed on, e.g., lamp posts, could shorten the transceiver distance. With the low height cells, however, the blockage probability increases.  Various blockage avoidance methods have been proposed for stationary and low mobility systems. For example, with context information the network  learns the environment and realizes proper resource allocation. The deployment of \ac{IAB} \cite{madapatha2020integrated}/relays, cooperative transmission \cite{fang2020hybrid} and \ac{NLoS} back-up links \cite{tunc2020} are also options for avoiding blockages.

\begin{figure*}
\centering
  \includegraphics[width=1.8\columnwidth]{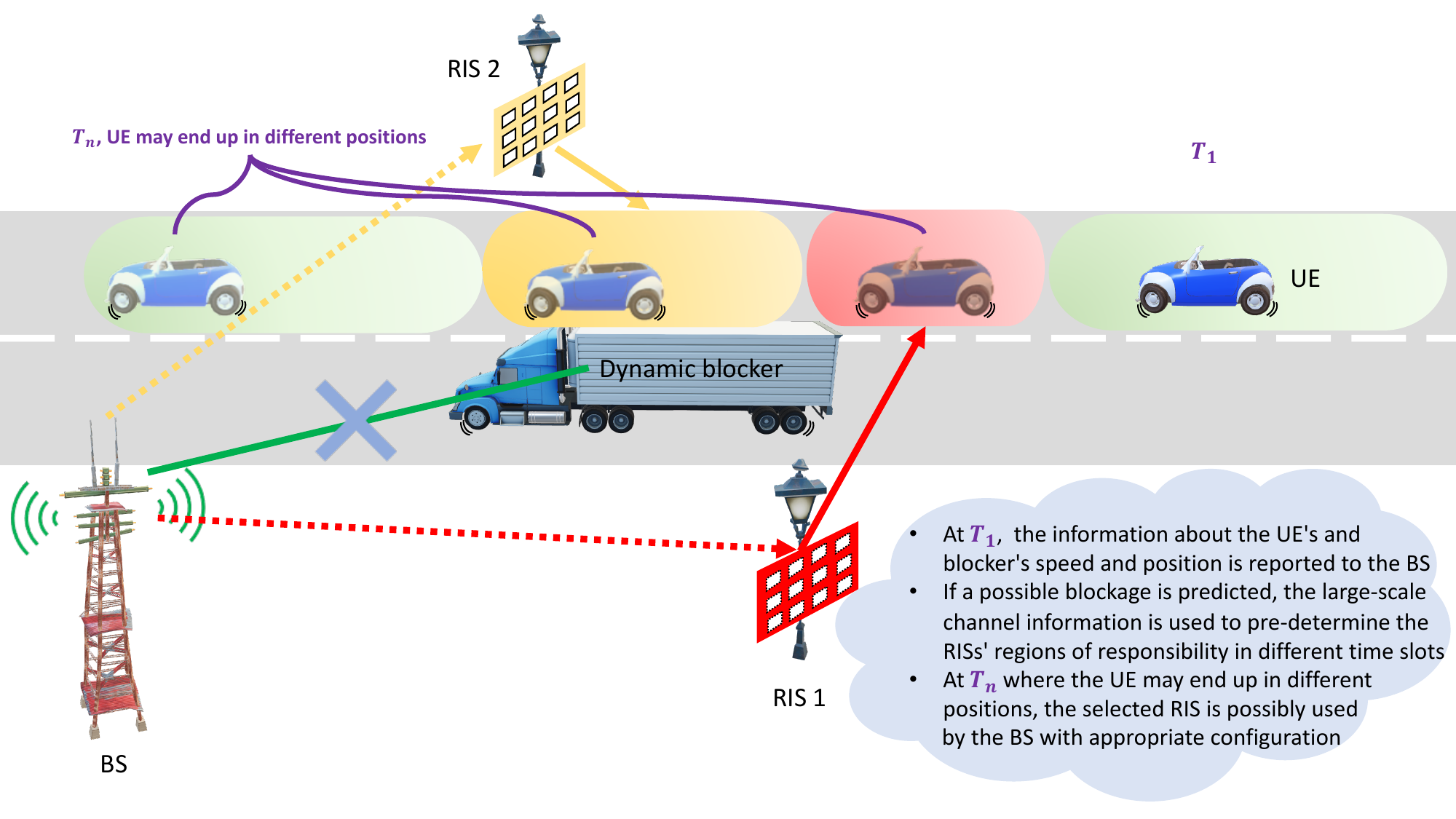}\\
\caption{An example of the proposed setup. At time $T_1$, the \ac{UE} and the blocker speed/position information is reported to the \ac{BS}. Then, the \ac{BS} exploits the information, along with the large-scale channel properties, to pre-select  the appropriate RIS to be used in Slot $T_n$. }
\label{fig_1}
\end{figure*}

Blockage  becomes more problematic for high mobility \ac{IoV} systems, e.g., in highways or rural areas. Here, due to the lack of good reflectors (e.g., buildings) and the mobility of vehicles, the \ac{NLoS} back-up links are rare and sustain for a short period which limit the effectiveness of \ac{NLoS} back-up links.  Recently,  \cite{guo2021iciot} proposes a dynamic \ac{BS} handover scheme for blockage avoidance  using additional large-scale predictor antennas  \cite{guo2021predictor} mounted on top of vehicles. However, deploying extra \acp{BS}/antennas  may not be commercially viable, and requires  backhauling/handover with possible failures/delays.  More importantly, deploying a \ac{BS}/\ac{IAB} along a highway/inter-city road may not be feasible if electricity connection is missing.

Instead of deploying additional \acp{BS}, one may consider the concept of \ac{RIS}, as a low-power alternative solution with, e.g., off-grid solar cell deployment. \ac{RIS} has recently   emerged as a cheap, low-power and flexible solution to  manipulate the wireless environment, resulting in better propagation channel. Particularly, with a proper deployment, \acp{RIS} can cover different areas of the highways/inter-city roads, which makes it possible to bypass dynamic blockages with no need for handover, backhauling,  and wired energy supply. However, while \ac{RIS} is of interest in stationary/low mobility networks, as explained in the following, it faces various challenges in terms of, e.g., initial access, \ac{CSIT} acquisition and/or imperfect beam reflection which may limit the \ac{RIS} efficiency at high speeds.  

In this article, we study the potentials and challenges of   \ac{RIS}-based  \ac{IoV} communications  in highways/inter-city roads with dynamic blockage pre-avoidance. We concentrate on two key challenges of \ac{RIS} in high-speed \ac{IoV} networks, namely, \ac{CSIT} acquisition and imperfect beam reflection. To reduce the \ac{CSIT} acquisition overhead, we propose a large-scale fading-based service region prediction scheme. Here, the service regions of different \acp{RIS} are learnt by the network beforehand for predicted dynamic blockages positions. In this way, without instantaneous \ac{CSIT}, the \ac{BS} can exploit the vehicle and the dynamic blockers speed information to pre-select the \acp{RIS} in different time slots. Then, we study the hardware aspects such as the transceiver impairments and phase noise effects leading to imperfect reflections, and compare the performance of the \ac{RIS}-based scheme with multiple candidate techniques.  

Our results reveal that with region prediction, our proposed \ac{RIS}-assisted scheme can reach  performance close to the cases with additional \acp{BS} utilizing handover as well as  \ac{RIS}- or network controlled repeater-assisted schemes with perfect CSIT, yet with low complexity. On the other hand,  the cost-efficiency trade-off of the \ac{RIS} may affect the usefulness of  \acp{RIS} in practice.

\section{Internet-of-Vehicles using RIS}

\ac{RIS} is composed of a 2D array of reflecting elements, where each element acts as a passive reconfigurable scatter that can be programmed to change an impinging  wave in a controlled way. The elements are low-cost passive surfaces with no need for dedicated power sources, and the radio waves impinged upon them is forwarded without the need of employing power amplifier/RF chain. As a result, \ac{RIS}  requires only low-rate control link, low energy supply, and no backhaul connections. 

One of the main applications of \ac{RIS} is to remove blind spots and provide the \acp{UE} with  alternative links when the direct \ac{UE}-\ac{BS} link experiences poor channel quality due to, e.g., blockage \cite{wu2020ICtowards,zhou2021TVTsto}.  This is specially of interest in \ac{mmw} communication, as the \ac{mmw} signal suffers from high penetration loss/low diffraction from objects.   

To integrate the \ac{RIS} into the network and bypass the blockage, the beam patterns of the \ac{BS} and the \ac{RIS} need to be jointly reconfigured. This, however, requires accurate \ac{CSIT} of both the \ac{BS}-\ac{RIS} and the \ac{RIS}-\ac{UE} links as well as accurate positioning. To acquire the \ac{CSIT}, one can turn the \ac{RIS} into the absorbing mode (if RF chain is deployed) and apply conventional channel acquisition methods. Alternatively, without explicit \ac{CSIT}, the reflection coefficients of the \ac{RIS} can still be optimized from the feedback of the \acp{UE}. For example, with pre-defined codebook, the \ac{BS} and the \ac{RIS} can sweep through all  beam patterns and iteratively obtain the best option. Such an additional overhead, compared to the direct \ac{BS}-\ac{UE} link \ac{CSIT} acquisition/beamforming planning, may be acceptable in the cases with stationary \acp{UE}, as the \ac{CSIT} acquisition and beamforming update can be performed during the network planning with no need for frequent updates. The problem, however, becomes challenging in high-speed  \ac{IoV} communications; With \ac{IoV}, both the vehicular \acp{UE} and the blockers move quickly, which affects the \ac{CSIT} accuracy of the  \ac{BS}-\ac{RIS}-\ac{UE} link.  The problem becomes even more challenging in, e.g., highways and rural areas, where to guarantee high reliability at high-speeds, one may require multiple \acp{RIS} and, depending on the \acp{UE}/blockers position, select the best \ac{RIS}-assisted path based on the instantaneous channel quality of the links/UEs’ positions. This is important because in a road scenario it is preferred to have a precise reflection pattern resulting in long beams covering the road over a long distance. Consequently, proper \ac{RIS} deployment will be critical to minimize beam tracking. However, as we explain in the following, present \ac{RIS} designs result in fuzzy and thereby short reflection pattern and, as a result, multiple \acp{RIS} are required to cover the road increasing beam tracking complexity.


Recent works \cite{zhou2021TVTsto,ozcan2021TVTrecon} study \ac{RIS}-assisted \ac{V2X} communication  in highways. Particularly, \cite{zhou2021TVTsto} concentrates on  beamforming optimization in the presence of random blockages, assuming perfect \ac{CSIT}. Then,  \cite{ozcan2021TVTrecon} investigates the optimal deployment of the \ac{RIS} in highway taking both the size and the operating mode of the \acp{RIS}  into account without explicit study on \ac{CSIT} acquisition.

\section{RIS-assisted Dynamic Blockage Pre-avoidance}

To enable multi-\ac{RIS} \ac{IoV}  communications, one needs to reduce the \ac{RIS} selection and configuration overhead as well as the sensitivity to the vehicles speed. For this reason, we propose a \ac{LSRPA} scheme in which the \acp{UE} and the blockers speed/position information is utilized along with the large-scale channel properties to predict and pre-select  the \ac{RIS} of interest, among multiple ones.

Consider the cases with either a macro or a small \ac{BS} along a highway/inter-city road, as illustrated in Fig. \ref{fig_1}. With network planning, the \ac{BS} location is normally optimized such that it covers a wide area of the road and static blockages are preferably avoided. However, dynamic blockages, due to, e.g., trucks, buses, are not encountered during the network planning and affect the achievable rate, specially in \ac{mmw} bands.   As an inexpensive and low-power solution to avoid dynamic blockages, specially if power supply is not available, \acp{RIS} with, e.g., off-grid solar cell deployment, can be installed along the road which will provide back-up links to the vehicular \acp{UE} when required.

With multiple \acp{RIS}, the main problem is to perform beam management in the \ac{BS} and each \ac{RIS} and select the best path. In the optimal case, one needs to know the instantaneous \ac{CSIT} of all paths for joint beamforming optimization and RIS selection \cite{zhou2021TVTsto,zappone2021TWCoverhead}. This, however, not only increases the \ac{CSIT} acquisition overhead, but also is not feasible at moderate/high speeds due to the \textit{channel aging} effect where the \ac{CSIT} acquired at a position soon becomes outdated due to the high mobility of the \ac{UE}/blocker.

With this background, the \ac{LSRPA} scheme follows the following procedure. At time slot $T_1,$ if the vehicular \ac{UE} detects a dynamic blockage, e.g., by a truck, it estimates the speed and the position of the blocker, e.g., using  cameras, lidars. Then, along with its own speed/position information, the \ac{UE} informs the \ac{BS} about the speed and the position of the dynamic blocker (As an alternative approach, each vehicle can inform the \ac{BS} about it own speed/position information). Knowing the blocker speed/position information at $T_1$, the \ac{BS} predicts the blocker position at Slot $T_n$. Then, the \ac{BS} utilizes the large-scale channel condition, i.e., the average performance which has been learned over time for the different blocker positions, to find the appropriate regions of interest to be covered by different \acp{RIS} in different time slots. Then, the \ac{BS} exploits the \ac{UE} speed/position information provided at $T_1$ to predict the \ac{UE} position at Slot $T_n$ and pre-select the appropriate path towards the \ac{UE}, either through direct \ac{BS}-\ac{UE} connection or via an \ac{RIS}-assisted link. Finally, at Slot $T_n$, only the instantaneous \ac{CSIT} of the pre-selected path, and not all possible paths, is acquired and the \ac{BS}/\ac{RIS} beamforming is adapted accordingly.

In this way, the \ac{LSRPA} setup reduces the \ac{CSIT} overhead/channel aging effect, and makes it possible to provide the vehicular \acp{UE} with fairly constant \ac{QoS}. Note that, to determine the regions, one does not need  to have extremely accurate information about the  speed and position information of the \ac{UE}/blocker. Such information is well achievable in, e.g., car platooning, connected vehicle or cruise control setups. Specially, in highways/inter-city roads,  slow large vehicles (resp. high-speed vehicles) travel typically in the outermost (resp. innermost) lanes with predictable trajectory, which simplifies the positioning. Thus, with a fairly predictable vehicles mobility, a limited number of beam transitions may be required and, consequently, e.g., an AI-based blockage prediction scheme can well reduce the \ac{RIS} selection/coordination overhead.

\begin{figure}
\centering
  \includegraphics[width=1.0\columnwidth]{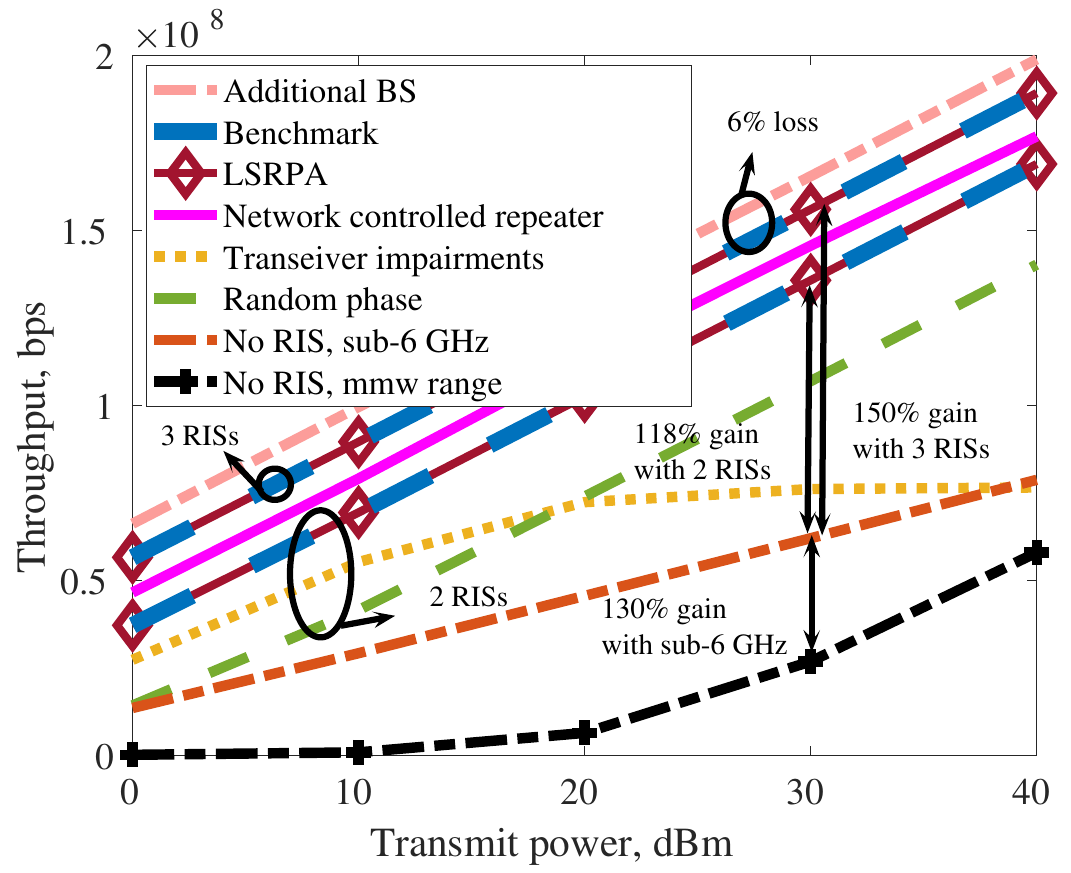}\\
\caption{Throughput as a function of the \ac{BS} transmit power with dynamic blockage. The \ac{BS} and the \ac{UE} are  equipped with 16 and 4 antennas, respectively. The \acp{RIS} have 200 elements.  The transmit power at the repeaters is set to 32 dBm. With transceiver impairment,  the proportionality coefficients which describe the severity of the distortion noises at the transmitter and the receiver are set to $0.05^2$ (see \cite{zhe2021twcachievable} for the details of the hardware impairment model). The \ac{BS}-\ac{RIS} 1/repeater 1 and \ac{BS}-\ac{RIS} 2/repeater 2  hop distances are set to 200 m and 126 m, respectively. The hop distance between the \ac{BS} and the additional \ac{BS} and \ac{RIS} 3 are 1500 m and 150 m, respectively. Finally, the results of sub-6 GHz refer to the case that the \ac{BS} switches to sub-6 GHz with  no \ac{RIS}. Except for the results of sub-6 GHz with 20 dB \ac{VPL}, 2.8 GHz frequency and 5 MHz bandwidth, the rest of the results are obtained  with 40 dB \ac{VPL}, 28 GHz frequency and 10 MHz bandwidth.}
\label{fig_2}
\end{figure}

\begin{figure}
\centering
  \includegraphics[width=1.0\columnwidth]{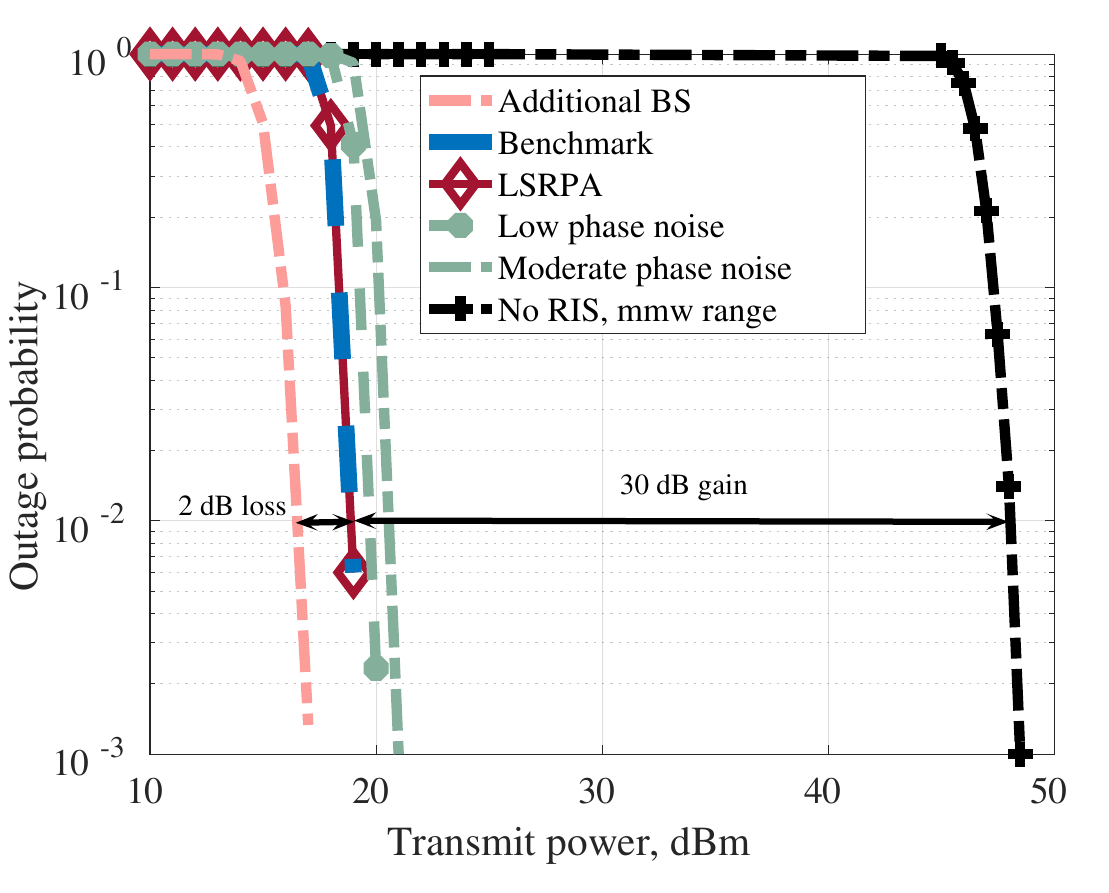}\\
\caption{The outage probability versus the \ac{BS} transmit power with dynamic blockage and two \acp{RIS}. The number of antennas for the \ac{BS} and the \ac{UE} are 16 and 4, respectively, 100 elements at each \ac{RIS} and 40 dB \ac{VPL} at 28 GHz. The threshold of outage is set to 8 bps. The hop distance between the \ac{BS} and the additional \ac{BS} is 5000 m. The \ac{BS}-\ac{RIS} 1 and \ac{BS}-\ac{RIS} 2  hop distances are set to 200 m and 126 m, respectively.}
\label{fig_3}
\end{figure}

\begin{figure}
\centering
  \includegraphics[width=1.0\columnwidth]{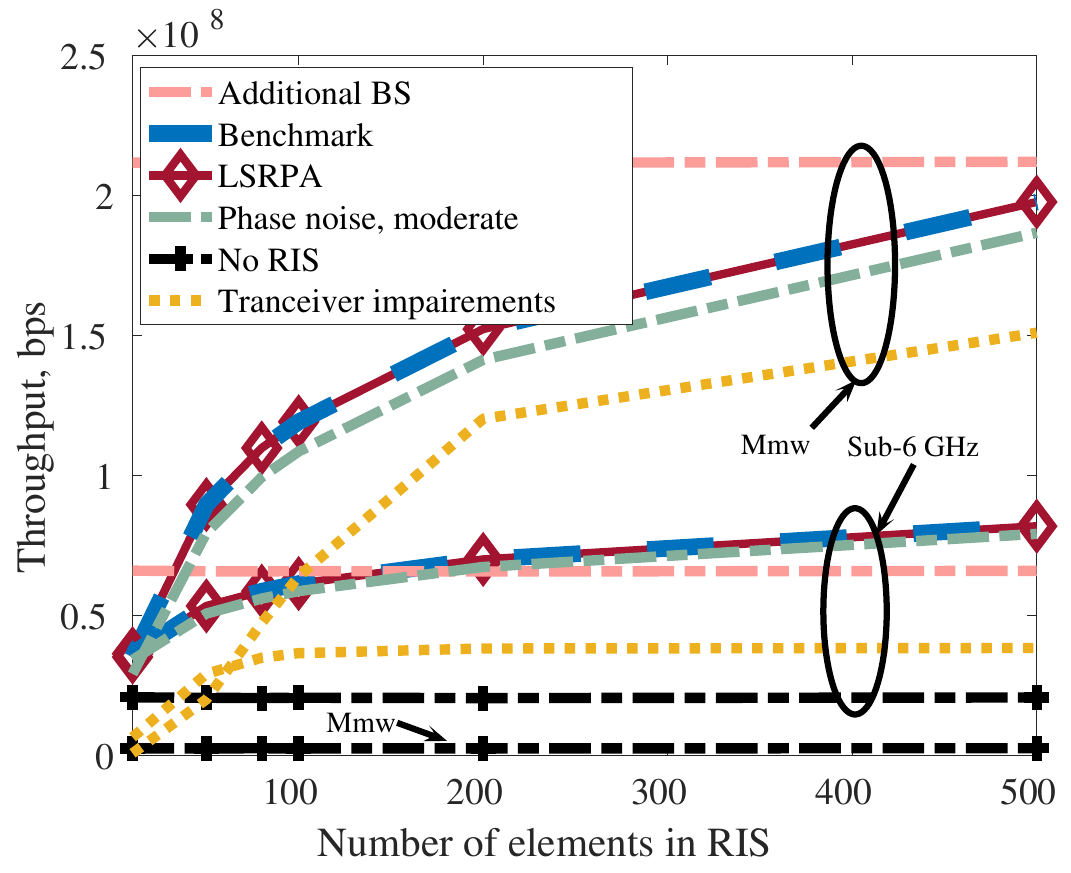}\\
\caption{Throughput as a function of the number of  elements in the \acp{RIS} with dynamic blockage. The bandwidth for \ac{mmw} (28 GHz, 16-antenna \ac{BS}) and sub-6 GHz (2.8 GHz, 8-antenna \ac{BS}) are set to 20 MHz and 5 MHz, respectively. Here, transmit power is 15 dBm and \ac{VPL} is set to 40 dB for 28 GHz and 25 dB for 2.8 GHz. The \ac{BS}-\ac{RIS} 1 and \ac{BS}-\ac{RIS} 2  hop distances are set to 200 m and 126 m, respectively.}
\label{fig_6}
\end{figure}

Considering the cases with one \ac{BS} and two or three \acp{RIS} with (im)perfect reflection efficiency, in the following, we evaluate the performance of the \ac{LSRPA} scheme, in comparison with other alternative techniques. We use typical \ac{RIS} setups as in, e.g., \cite{zappone2021TWCoverhead}, and the \ac{RIS} beamforming is performed using  \cite[Algorithm 1]{zappone2021TWCoverhead}. As the metric of interest, we consider the outage probability and the throughput where the throughput is defined as the total number of successfully decoded bits per total transmission delay. We consider both sub-6 GHz (2.8 GHz) and \ac{mmw} (28 GHz) bands and different numbers of \ac{RIS} elements (10-500), with the details of the parameter settings  given in the figures captions. In Figs. \ref{fig_2}, \ref{fig_3} and \ref{fig_6} the effect of the hardware impairments are studied, otherwise ideal \ac{RIS} is considered. To evaluate the efficiency of the \ac{LSRPA} scheme, we compare its performance with the following alternative schemes:
\begin{itemize}
    \item\textit{Additional \ac{BS}.} Here, instead of \acp{RIS}, an additional \ac{BS}, considerably more expensive than an \ac{RIS}  and requiring backhaul, is added to the network, and cooperative handovers are used to bypass the blockage \cite{guo2021iciot}. Note that, compared to the \ac{RIS} setup, the deployment of \ac{BS} is less intensive and it normally has larger coverage area.
    \item\textit{Network controlled repeater.} Network controlled repeater is a normal repeater with beamforming capabilities which forwards the signal to the destination using amplify-and-forward relaying. That is, compared to RIS, network controlled repeater is a more advanced and expensive node with more focused beamforming capability/accuracy and active signal amplification.
    \item\textit{Benchmark.} A genie-aided scheme using \acp{RIS} where, in each time slot, the \ac{BS} utilizes the instantaneous \ac{CSIT} of all links to search over all, possibly \ac{RIS}-assisted, paths with optimal beam management at the \ac{BS}/\ac{RIS}, and selects the one with the best performance. 
    \item\textit{Random phase.} Here, while we follow the same approach as in the \ac{LSRPA} scheme to select the appropriate \ac{RIS}, the phase matrix of the selected \ac{RIS} is not optimized and, instead, is considered to be random. The setup is of interest in the cases where either the instantaneous \ac{CSIT} of the selected path is not available or the \ac{RIS} has no adaptation capability.  
    \item\textit{No \ac{RIS}.} The \ac{UE} is served by direct \ac{BS} communication,  either at low or \ac{mmw} band, at the risk of possible blockage.  Here, we model the blockage by the \ac{VPL} which is set to 20 dB in sub-6 GHz (e.g., 2.8 GHz) and 40 dB in \ac{mmw} frequencies (e.g., 28 GHz). 
\end{itemize}

As expected, without \ac{RIS}/network controlled repeater/cooperative \acp{BS}, if the \ac{BS} switches to sub-6 GHz when blockage occurs, better throughput is observed compared to the cases with \ac{mmw} communications, thanks to lower \ac{VPL} at low frequencies (Fig. \ref{fig_2}). However, compared to the cases with no \ac{RIS}, the implementation of the \acp{RIS} improves both the coverage and the throughout significantly. For instance, considering the parameter settings of Fig. \ref{fig_2} and the \ac{BS} transmit power 30 dBm, \ac{RIS}-assisted communication in \ac{mmw} range increases the throughput, compared to no-\ac{RIS} scenario  operating in 2.8 GHz, by 118\% and 150\%, in the cases with two and three \acp{RIS}, respectively. Moreover, as shown in Fig. \ref{fig_3}, with an outage probability $10^{-2}$, bypassing dynamic blockages by two \acp{RIS} leads to 30 dB gain of the transmit power.

Importantly, our proposed \ac{LSRPA} scheme shows the same performance as in the genie-aided benchmark approach with perfect instantaneous \ac{CSIT} of all paths, and no coverage/throughput loss is observed by \ac{RIS} selection based on large-scale fading. This leads to considerably lower overhead, specially as the number of \acp{RIS} increases, and reduces the channel aging effect. Interestingly, the performance degradation due to the implementation of the \ac{RIS} instead of cooperative \acp{BS} is negligible. For instance,   with the \ac{BS} transmit power 30 dBm (resp. outage probability $10^{-2}$), only $18\%$ throughput reduction (resp. 2 dB power increment) is observed, compared to the cases with cooperative \acp{BS}, when the network is equipped with two \acp{RIS} (Figs. \ref{fig_2}-\ref{fig_3}). Also, with three \acp{RIS} the relative throughput loss, compared to the cases with additive \acp{BS}, reduces to only $6\%$.  This is important because not only the additional \ac{BS} installation leads to considerable cost/energy consumption increment, but it also is at the cost of backhauling and handover with possible delays/failures. Finally, as expected, the implementation of network controlled repeater leads to slightly higher throughput compared to \acp{RIS} since, along with phase adaptation, network controlled repeater is capable of power amplification also. However, compared to RIS, network controlled repeater is a relatively more expensive node requiring dedicated power supply.  In this way, the proposed \ac{LSRPA} scheme can be considered as a cheap solution for high-rate uninterrupted \ac{IoV} communications in highways/rural areas with low signaling overhead, especially when electricity connection is not available.

\begin{figure}
\centering
  \includegraphics[width=1.0\columnwidth]{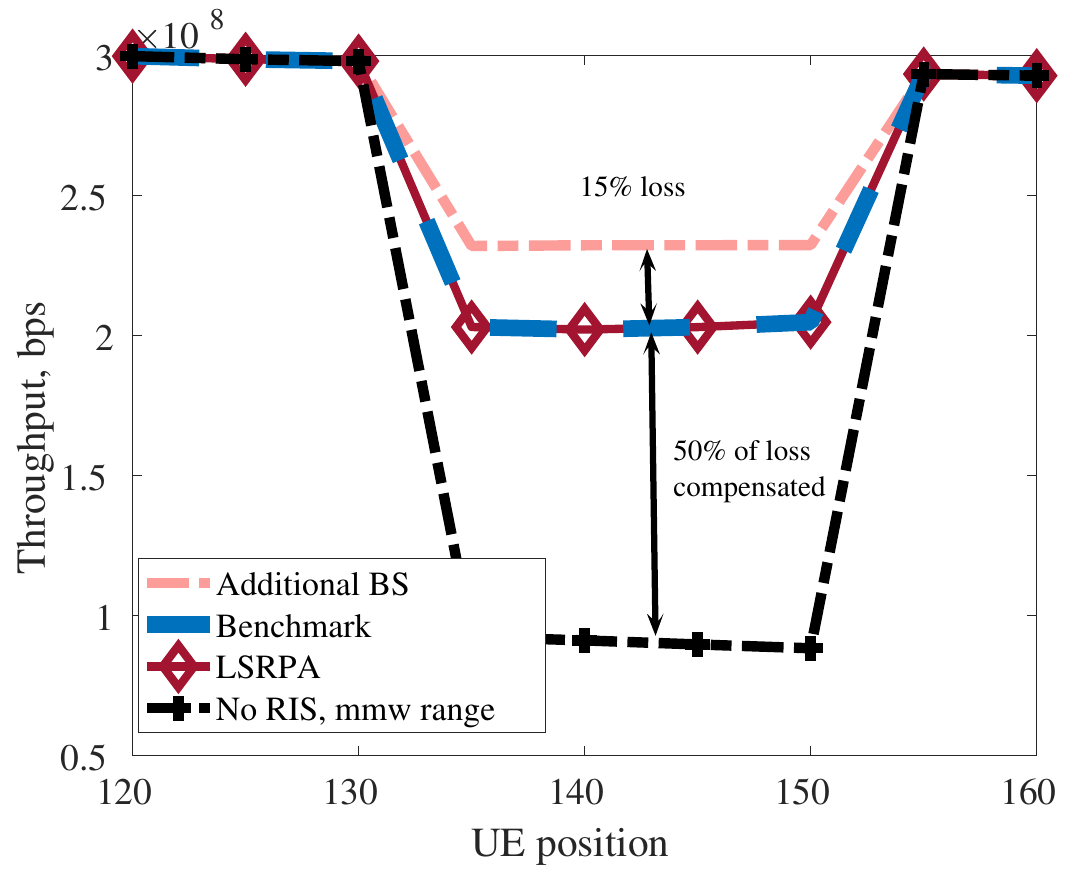}\\
\caption{An example of the proposed scheme when blockage occurs, 16/4 antennas at the \ac{BS}/\ac{UE}, 50 dBm transmit power, 40 dB \ac{VPL} at 28 GHz, and 200 elements in the \acp{RIS}. The hop distance between the \ac{BS} and the additional \ac{BS} is 1500 m. The \ac{BS}-\ac{RIS} 1 and \ac{BS}-\ac{RIS} 2  hop distances are set to 200 m and 126 m, respectively.}
\label{fig_4}
\end{figure}

\begin{figure*}
\centering
  \includegraphics[width=1.8\columnwidth]{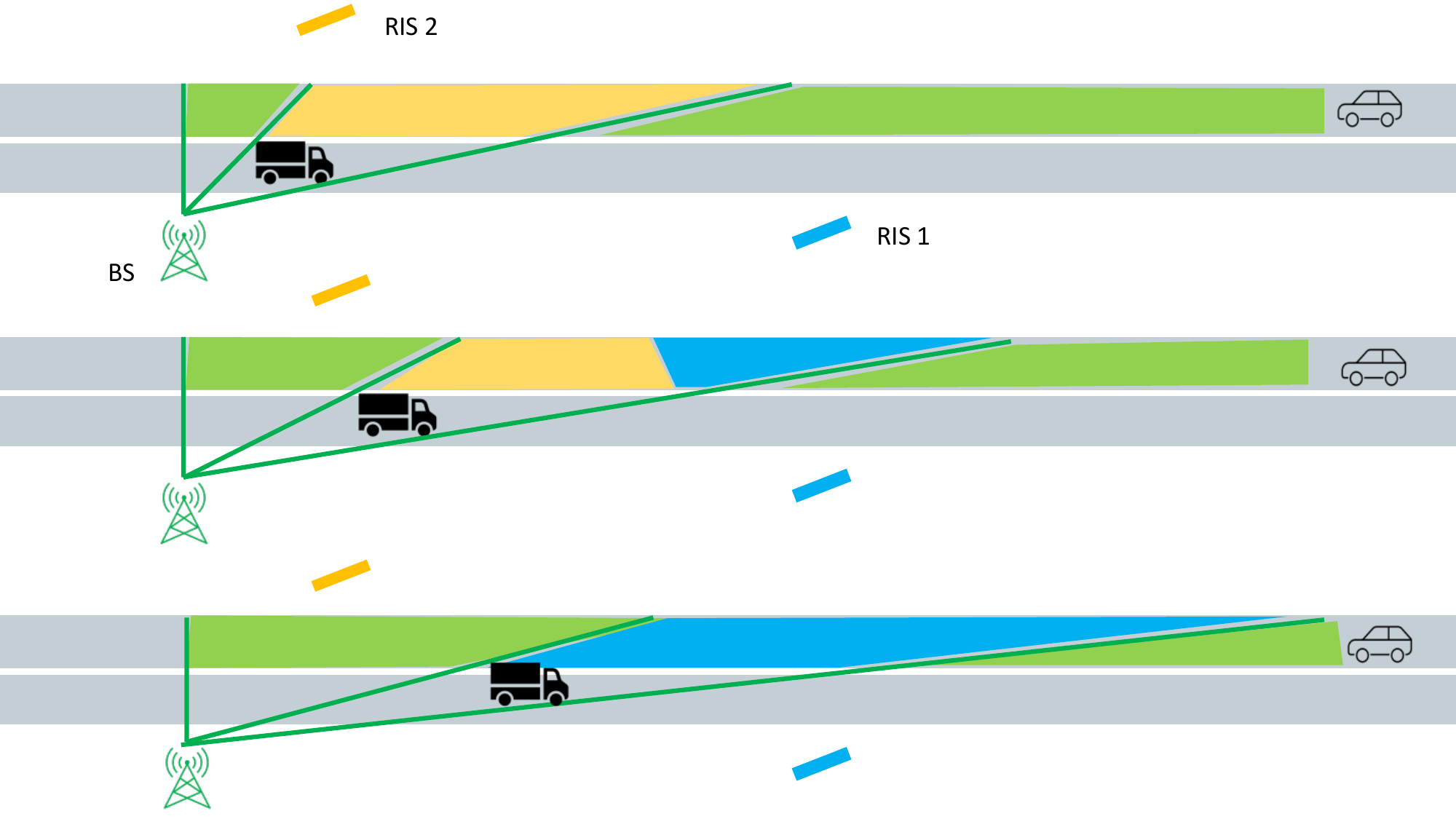}\\
\caption{\acp{RIS} service regions for different positions of the dynamic blocker. The green regions represent the areas with direct \ac{BS}-\ac{UE} communications. The blue and the yellow regions are the regions of responsibility for \ac{RIS} 1 and \ac{RIS} 2, respectively.}
\label{fig_5}
\end{figure*}

To demonstrate the efficiency of the \ac{LSRPA} scheme, in Fig. \ref{fig_4} we present the \ac{UE} throughput in an example scenario, with possible dynamic blockages. Moreover, Fig. \ref{fig_5} shows the \ac{BS}'s and the \acp{RIS}' regions of responsibility, determined based on the large-scale fading, for different positions of the dynamic blocker. As shown, with high frequencies, dynamic blockage deteriorates the system performance significantly. However, with blockage pre-avoidance, either through \acp{RIS} or cooperative \acp{BS}, 50\% of the throughput loss is compensated in this example scenario (Fig. \ref{fig_4}). Without blockage, the \ac{UE} is preferably served through the direct \ac{BS}-\ac{UE} link, as the reflection losses of the \acp{RIS} are avoided. With blockage, however, the road is partitioned into various regions for which different \acp{RIS} are responsible, and the \ac{UE} is served with the closest RIS experiencing better long-term channel quality (Fig. \ref{fig_5}). The \acp{RIS}' regions of responsibility are adapted dynamically with the blocker position.

In Fig. \ref{fig_6}, we study the effect of different parameters on the system throughput. In general, \ac{mmw}-based communication outperforms that in the sub-6 GHz with more bandwidth resources, if the blockage is avoided via \acp{RIS}/cooperative \acp{BS}. This is not the case if the blockage is not bypassed. Finally, for a given number of antennas/transmit power at the cooperative \ac{BS},  increasing the number of \ac{RIS} elements may end up in better throughput, at the cost of \ac{CSIT} acquisition overhead and \ac{RIS} cost increment.


\section{Towards RIS Practicality}

Although the simulation results show great potentials for \ac{RIS}-assisted communications, there are key issues to be solved before \acp{RIS} can be used in practice. In the following, we elaborate on some key challenges. 

\subsection{Cost-efficiency Trade-off}

One of the main motivations of using \ac{RIS}, and not small access points such as relays, network controlled repeaters, and \acp{IAB}, is the cost and energy reduction. With a cheap node, however, hardware imperfection may affect the reflection quality of the \acp{RIS}. Particularly, it is likely that, in practice, \acp{RIS} may provide an imperfect reflection because of, e.g.,  transceiver impairments and phase noise. To study the effect of hardware impairment, in Figs. \ref{fig_2}, \ref{fig_3}, and \ref{fig_6}, we consider two different imperfection models:
\begin{itemize}
    \item\textit{Transceiver impairments.} The performance of the \ac{RIS}-assisted system could be affected by  a mismatch between the intended transmitted/received signal and the actual transmitted/received signal. Such a mismatch can be  modeled as inherent hardware impairments caused by, e.g., non-linear amplifier and quantization error \cite{zhe2021twcachievable}. Specifically, the impairment at the \ac{RIS} side can be modeled as uniformly distributed random variables, while at the transceiver side  the additive distortion noise as well as the phase drift and the thermal noise are used to describe the hardware imperfection \cite{zhe2021twcachievable}. In this way, the ergodic capacity of the system can be bounded as a function of the transceiver  proportionality coefficients that describe the severity of the distortion as well as the \ac{SNR} \cite[Theorem 1]{zhe2021twcachievable}.
    \item\textit{Phase noise.}  Another hardware limitation is  when the channel cause additional phase deviation and it is not estimated properly at the receiver side. The phase noise can also be caused by the discrete set of phases. Here,  the phase noise can be modeled as a uniform distribution over certain region (determined by, e.g., estimation accuracy and quantization phases) which is symmetric around zero with mean zero \cite{badiu2020WCL}.
\end{itemize}

As shown in Figs. \ref{fig_3} and \ref{fig_6}, the effect of phase noise on the outage probability and the throughput is negligible, as long as the phase noise is not considerably high. However, the effect of phase noise increases slightly at \ac{mmw} range (Fig. \ref{fig_6}). Also, as illustrated in Fig. \ref{fig_2}, the effect of transceiver impairment on the throughput is negligible at low \acp{SNR}. At moderate/high \acp{SNR}, however, which is the range of interest, transceiver impairment reduces the efficiency of the \acp{RIS} significantly,  where even higher throughput may be achieved by direct communication of the \ac{BS} operating at sub-6GHz compared to the cases with \ac{RIS}-assisted communication in \ac{mmw} range (Fig. \ref{fig_2}). Moreover, to reach the same performance as in the cases with cooperative \acp{BS}, significantly larger number of \ac{RIS} elements are required, if realistic conditions with transceiver impairments are taken into account (Fig. \ref{fig_6}). This, in turn, increases not only the implementation cost but also the \ac{RIS} management overhead.

In this way, there is a cost-efficiency trade-off where with low cost the poor performance of the RIS may affect its efficiency while with an expensive \ac{RIS} other alternative technologies such as \acp{IAB} and network controlled repeaters may be better choices with higher capabilities. Thus, the practical implementation of the \ac{RIS} still requires further justifications. 

\subsection{Standardization and Network-level Performance}

\ac{RIS} has not yet been considered by 3GPP in  Release 17 and before. During the initial discussions on 3GPP Release 18, \ac{RIS}-assisted communication was suggested by some companies as a possible technology to be considered in Release 18  study-item on network controlled repeaters, e.g. \cite{3gppris2}. However, it was decided not to continue with it, and to leave it for possible use in beyond 5G. 

Without standardization, the integration and utilization of \acp{RIS} may be challenging, as it may affect the performance of the rest of the network. Importantly,  \ac{RIS} networks may suffer from an increased
network interference, specially in the cases with fuzzy
reflections of the \acp{RIS}. Such an interference may possibly also affect the adjacent channels which is a potential show stopper for the \ac{RIS} implementation,  at least in present day \ac{RIS} implementations. Furthermore, if \ac{RIS} performance is not specified, it may still reflect the incoming signals in different directions that are not desired.  For these reasons, there is a need for standardized mechanisms where, while the \ac{RIS}-assisted communication improves the experienced \ac{QoS} of the intended receiver, reliable network-level performance is guaranteed. 

Along with cost-efficiency trade-off, standardization and interference management, there are still other issues to be investigated in \ac{RIS} networks. For instance, although there are few practical
evaluations, e.g., \cite{geo2021design}, multiple testbed evaluations are required to validate the efficiency of the \ac{RIS}, in competition to alternative
technologies. Here, the practicality and the propagation model of the \ac{RIS} still require testbed evaluations specially in high frequencies and \ac{MIMO} setups. Finally, the relative cost reduction of the \acp{RIS}, compared to, e.g., network controlled repeaters and \acp{IAB}, is not yet clear as a large part of, e.g., site rental and installation, costs still remain in \acp{RIS}. This calls for realistic cost analysis of the \ac{RIS} networks. 

In summary, the selection of the best approach depends on different parameters such as backhaul availability, implementation cost, availability of electricity connection and the deployment. However, while more advanced nodes such as network controlled repeaters and \acp{IAB} seem to be more appropriate candidates guaranteeing proper interference management/network-level performance, \ac{RIS} may offer a low-power alternative with, e.g., off-grid solar cell deployment if electricity connection is not available.  Here, \ac{RIS} could provide fast deployment to improve the network coverage/reliability until/if electricity connection is provided.


\section{Conclusions}
We studied the potentials and challenges of RIS-assisted communication for blockage pre-avoidance in moving networks. As we showed, RIS pre-selection and blockage prediction gives the chance to robustify the network performance against dynamic blockages with an acceptable CSI acquisition overhead. However, there are still various issues such as hardware impairment, cost-efficiency trade-off, interference management, standardization  and performance improvement, to be competitive with alternative technologies, which should be solved before RIS can be practically implemented  in IoV networks.

\section*{Acknowledgement}
This work was supported by the European Commission through the H2020 Project Hexa-X under Grant 101015956, and in part by the Gigahertz-ChaseOn Bridge Center at Chalmers in a project financed by Chalmers, Ericsson, and Qamcom.

\bibliographystyle{IEEEtran}
\bibliography{main.bib}

\begin{IEEEbiography}{Hao Guo} [S'17] (hao.guo@chalmers.se)  is currently pursuing his PhD degree with Department of Electrical Engineering, Chalmers, Sweden.
\end{IEEEbiography}

\begin{IEEEbiography}{Behrooz Makki} [M'19, SM'19] works as a Senior Researcher in Ericsson Research, Sweden. 
\end{IEEEbiography}

\begin{IEEEbiography}{Magnus Åström}  works as a Research Leader in Ericsson Research, Sweden. 
\end{IEEEbiography}

\begin{IEEEbiography}{Mohamed-Slim Alouini} 
[S'94, M'98, SM'03, F'09]  is a Distinguished Professor of Electrical Engineering in King Abdullah University of Science and Technology, Saudi Arabia. 
\end{IEEEbiography}

\begin{IEEEbiography}{Tommy Svensson} [S’98, M’03, SM’10]  is a  Full Professor in Communication Systems at Chalmers University of Technology, Sweden.
\end{IEEEbiography}

\end{document}